\documentclass[11pt,a4paper]{article}
\usepackage{jcappub}

\title{Chern-Simons anomaly as polarization effect }
\author[a,b,1]{V. B. Semikoz,%
\note{On leave from IZMIRAN.}}
\author[b]{J.~W.~F.~Valle}
\affiliation[a]{Pushkov Institute of the Terrestrial Magnetism, the Ionosphere\\
and Radiowave Propagation of the Russian Academy of Sciences (IZMIRAN)\\
Troitsk, Moscow region, 142190, Russia}
\affiliation[b]{AHEP Group, Institut de F\'{\i}sica Corpuscular --
  C.S.I.C./Universitat de Val{\`e}ncia \\
  Edificio Institutos de Paterna, Apt 22085, E--46071 Valencia, Spain,
}

\emailAdd{semikoz@ific.uv.es}
\emailAdd{valle@ific.uv.es}

\abstract{The parity violating Chern-Simons term in the epoch before the
  electroweak phase transition can be interpreted as a polarization
  effect associated to massless right-handed electrons (positrons) in
  the presence of a large-scale seed hypermagnetic field.
  We reconfirm the viability of a unified seed field scenario relating
  the cosmological baryon asymmetry and the origin of the
  protogalactic large-scale magnetic fields observed in astronomy.  }
 \arxivnumber{1104.3106}
\begin{document}
\maketitle

\section{Introduction}

Magnetic fields are known to play an important role in the physics of
a variety of astrophysical objects, from stars to galaxies and galaxy
clusters. The nature of the initial weak seed fields for the following
dynamo or turbulent amplification is largely unknown
\cite{Kulsrud:2007an,Kronberg:1993vk}. It might be that the seed
fields are produced during the epoch of galaxy formation from
frozen-in magnetic fields of proto-galaxies experiencing gravitational
collapse, or ejected by the first supernovae or active galactic
nuclei.

Primordial hypermagnetic fields may alternatively arise from phase
transitions in the very early universe, before electroweak phase
transition, such as during the inflationary
epoch~\cite{Grasso:2000wj}.

The clue for choosing between these possibilities may lie in
measurements of the initial seed fields. However, up to recently there
was little hope that extremely weak fields outside galaxies and galaxy
clusters would be ever be detected. Nevertheless, the extragalactic
magnetic fields originated from a seed magnetic field in the early
Universe could, in principle, be observed in $\gamma$ -astronomy with
help of satellite instruments like FERMI \cite{Neronov:2009gh}.

Here we adopt the second scenario with a cosmological seed field
present before the epoch of electroweak phase transition.
Starting from the effective Lagrangian for the hypercharge gauge field
in the presence of the seed field, we derive the parity violating
Chern-Simons term as resulting from a polarization effect of the seed
field upon the primeval plasma. This way we provide a clear physical
interpretation for the Chern-Simons term, as resulting from the
right-handed electron chemical potential and the associated asymmetry.

Solving Faraday equation one finds exponential amplification of the
seed hypermagnetic field~\cite{Semikoz:2007ti} through the
$\alpha^2$-dynamo mechanism. Such a seed hypermagnetic field is
subsequently converted into a seed Maxwellian field.
It has been shown in Ref.~\cite{Semikoz:2009ye} that, thanks to the
anomaly term~\cite{PhysRevD.14.3432,kuzmin:1985mm}, such seed field
may induce a sizeable baryon asymmetry of the Universe, providing an
alternative to conventional leptogenesis~\cite{buchmuller:2005eh}
which directly involves nonzero neutrino masses~\cite{Schwetz:2011qt},
induced by the seesaw mechanism~\cite{valle:2006vb}.

We revisit briefly such ``magneto-baryogenesis'' scenario and confirm
that, indeed, the required field strength estimates needed to account
for the cosmological baryon asymmetry match those inferred by current
galactic magnetic field observations, providing a remarkable
connection between astronomical and cosmological observations.

\section{Chern-Simons as Polarization}
\label{sec:polar-chern-simons}

Despite the various conserved charges of the standard model, the early
electroweak plasma at temperatures above the characteristic
chiral-flip temperature, $T_{RL} \sim$~few~10~TeV, is described by
just one non-zero chemical potential associated to right-handed
electrons, $\mu_{eR}$, with the corresponding number perturbatively
conserved~\cite{Giovannini:1997eg}.
Above $T_{RL}$ the statistically averaged standard model Lagrangian
density for the hypercharge field $Y_{\mu}$ contains a term 
\begin{equation}\label{right}
f_R(g^{'})<\bar{e}_R\gamma_{\mu}e_R>Y^{\mu}
\end{equation}
involving $\mu_{eR}\neq 0$.  Here $g^{'}$ is the $U_Y(1)$ gauge
coupling, and $f_R(g^{'})=g^{'}y_R/2$ plays the role of an "electric"
charge associated to $U_Y(1)$, $y_R=-2$ being hypercharge of the
right-handed electron.

For simplicity we take an external large-scale hypermagnetic field
${\bf B}_Y=(0,0,B_Y)$ directed along an arbitrary  "z" axis. 
In the presence such field one can take the statistical average using
the equilibrium density matrix for right-handed electrons (positrons),
\begin{equation}\label{Fermi}
f^{(e_R,\bar{e}_R)}_{\lambda^{'}\lambda}(\varepsilon(p_z,n, \lambda))=\frac{\delta_{\lambda^{'}\lambda}}{\exp [(\varepsilon(p_z,n, \lambda) \mp \mu_{eR})/T] + 1}~.
\end{equation}
The resulting Landau spectrum of the massless right-handed electrons
(positrons),
\begin{equation}\label{Landau}
\varepsilon(p_z,n, \lambda)=\sqrt{p_z^2+\mid f_R(g^{'})\mid B_Y(2n + 1 \mp \lambda )},
\end{equation}
depends on the Landau number $n=0,1,2,...$, and the spin projection on
the hypermagnetic field $\lambda=\pm 1$,
$(\sigma_z)_{\lambda^{'}\lambda}=\lambda
\delta_{\lambda^{'}\lambda}$. In Eqs. (\ref{Fermi}),(\ref{Landau}) the
upper sign applies to particles, the lower one to antiparticles. Note
that together with chirality $\gamma_5\Psi_{eR}=+\Psi_{eR}$ the spin
projection $\lambda$ is a good quantum number since
$[\gamma_5,\Sigma_z]=0$.

Thus, the corresponding macroscopic right-handed electron four-current
in eq.~(\ref{right}) includes the pseudovector term for which the
3-vector component ${\bf J}_5^Y$ is given by
\begin{eqnarray}\label{3vector}
&&J_{j5}^Y=\frac{f_R(g^{'})}{2}<\bar{e}\gamma_j\gamma_5e>=\frac{f_R(g^{'})}{2}
\sum_{n=0}^{\infty}\frac{\mid f_R(g^{'})\mid B_Y}{(2\pi)^2}\int_{-\infty}^{+\infty}{\rm d}p_z\times
\nonumber\\&&\times Tr\Bigl[\sigma_j\Bigl(f^{(e_R)}(\varepsilon(p_z,n,\lambda))-f^{(\bar{e}_R)}(\varepsilon(p_z,n,\lambda))\Bigr)\Bigr]=\nonumber\\
&&=\frac{f_R(g^{'})}{2}\sum_{n=0}^{\infty}\frac{\mid f_R(g^{'})\mid B_Y\delta_{jz}}{(2\pi)^2}\int_{-\infty}^{+\infty}{\rm d}p_z\sum_{\lambda}\lambda \Bigl[f^{(e_R)}_{\lambda\lambda}(\varepsilon(p_z,n,\lambda))-f^{(\bar{e}_R)}_{\lambda\lambda}(\varepsilon(p_z,n,\lambda))\Bigr]=\nonumber\\
&&=\frac{f_R(g^{'})\mid f_R(g^{'})\mid B_Y\delta_{jz}}{(2\pi)^2}
\int_0^{\infty}\left[\frac{1}{\exp \left[(p-\mu_{eR})/T\right] +1} -\frac{1}{\exp \left[(p+\mu_{eR})/T\right] +1}\right]{\rm d}p~.
\end{eqnarray}

Here summing over $\lambda$ and $n$ in the second line in
Eq. (\ref{3vector}) we used the cancellation of all degenerate terms
$n=1,2,...$. This happens separately for particles and for
antiparticles due to $\varepsilon_{n+1,1}=\varepsilon_{n, -1}$ and
$\varepsilon_{n+1, -1}=\varepsilon_{n,+1}$.
The asymmetry at the main Landau level in the last line gives exactly
$\mu_{eR}$ for integral, so that one obtains the mean pseudovector
current as
\begin{equation}\label{J5}{\bf J}_5^Y=-g^{'2}\mu_{eR}{\bf
    B}_Y/4\pi^2,\end{equation}
while its time component vanishes, $ J_{05}^Y=0$, due to the zero
global hypercharge condition $<Y>=0$.

As result of statistical averaging the effective standard model
Lagrangian density at finite fermion density $\mu_{eR}\neq 0$ in the
early hot plasma bath ($T>T_{EW}$) takes the
form~\cite{Giovannini:1997eg}:
 \begin{equation}\label{Lagrangian2}
L=-\frac{1}{4}Y_{\mu\nu}Y^{\mu\nu}  - J_{\mu}Y^{\mu} - \frac{g^{'2}\mu_{eR}}{4\pi^2}{\bf B}_Y{\bf Y},
\end{equation}
where $J_{\mu}$ is the vector (ohmic) current with zero time component
due to the electro-neutrality of the plasma as a whole, $J_0=<Q>=0$.

This way the physical meaning of the mean pseudovector current
eq.~(\ref{J5}) emerges, in terms of comoving right-handed electrons
and positrons at the main Landau level in the external hypermagnetic
field, in a way similar to the discussion given in
Ref.~\cite{Semikoz:2007ti}~\footnote{By contrast, we recall that the
  conventional derivation of the Chern-Simons
  term~\cite{Redlich:1984md,Laine:1999zi} involves the use of
  alternative one-loop diagrammatic calculations in finite temperature
  field theory.  }.
Thanks to the Coulomb force electrons and positrons move in the same
direction with respect to the hypermagnetic field, with a net electric
current ${\bf J}^Y_5$ resulting from a slight difference in their
densities if $\mu_{eR}\neq 0$.
In other words, the origin of the Chern-Simons interaction term ${\bf
  J}_5{\bf Y}$ as a polarization effect is a direct consequence of the
spin paramagnetism of electrons and positrons at the main Landau level
which leads to a magnetization in opposite directions, $\lambda=\mp
1$, weighted with different populations due to the asymmetry density,
$\mu_{eR}\neq 0$.  According to Faraday equation the current in
eq.~(\ref{J5}) induces a "wrong" {\it transversal} component of
hypermagnetic field $\nabla\times \alpha_Y{\bf B}_Y$, where
$\alpha_Y=-g^{'2}\mu_{eR}/4\pi^2\sigma_{cond}$ is the hypermagnetic
helicity parameter, $\sigma_{cond}\simeq 100 T$ is the plasma
conductivity.
This hypermagnetic field component winds around the pseudovector
current ${\bf J}_5^Y$ parallel to the self-consistent ${\bf
  B}_Y$. Note that such term violates parity, or the total
hypermagnetic field has both a vector and an axial vector components.

 \section{Magneto-baryogenesis revisited}

 We now briefly revisit the ``magneto-baryogenesis'' scenario proposed
 in Ref.~\cite{Semikoz:2009ye}. Its basic ingredient is a primordial
 hypercharge field that induces a nonzero lepton asymmetry of the
 early universe plasma through the Abelian anomaly
 for right electrons,
 \begin{equation}\label{Abelian}
 \partial_{\mu}j^{\mu}_{eR}=-\frac{g^{'2}y_R^2}{64\pi^2}Y_{\mu\nu}\tilde{Y}^{\mu\nu}.
 \end{equation}
 Irrespective of such anomaly one can re-derive Faraday equation from
 the effective Lagrangian we have obtained through the statistical
 averaging of the standard model Lagrangian in vacuum quantum field
 theory.  Following Ref.~\cite{Semikoz:2007ti}, we take the rest frame
 of the Universe as a whole, and re-obtain the $\alpha^2$-dynamo
 amplification of the primordial seed hypermagnetic field, instead of
 $\alpha\Omega$-dynamo mechanism of standard
 magnetohydrodynamics~\cite{1983flma....3.....Z}, namely

\begin{eqnarray}\label{arbitrary}
   &&B_Y(t)=B_0^Y\exp \left[\left(\frac{1}{\kappa} - \frac{1}{\kappa^2}\right)\int_{t_0}^{t}\frac{\alpha_Y^2(t^{'})}{\eta_Y(t^{'})}\right]=\nonumber\\
   =&&B_0^Y\exp \left[83\left(\frac{1}{\kappa} -
       \frac{1}{\kappa^2}\right)\int_x^{x_0}\frac{{\rm
         d}x^{'}}{x^{'2}}
     \left(\frac{\xi_{eR}(x^{'})}{0.0001}\right)^2\right].\nonumber\\
 \end{eqnarray}
 Here $\Lambda$ denotes an arbitrary scale of the hypermagnetic field
 $\Lambda=\kappa \eta_Y/\alpha_Y$, $\kappa>1$, where
 $\eta_Y=1/\sigma_{cond}$ is the magnetic diffusion coefficient;
 $\xi_{eR}=\mu_{eR}/T$ is the dimensionless right electron asymmetry
 parameter; $x=T/T_{EW}\geq 1$; the moment $x_0\gg 1$ corresponds to
 the initial time $t_0/t_{EW}=x_0^{-2}=(T_{EW}/T_0)^2$ when a tiny
 seed hypermagnetic field $B_0^Y$ starts to polarize hot plasma at the
 initial temperature $T_0\gg T_{EW}$.

 Note that the ``slope'' of hypermagnetic field enhancement in the
 $\alpha^2$ -dynamo mechanism given in eq.~(\ref{arbitrary}) differs
 from that obtained in Ref.~\cite{Semikoz:2007ti} using the net
 neutrino asymmetry instead of the right-handed electrons asymmetry
 used above, and dictated by the correct equilibrium conditions found
 in Ref.~\cite{Giovannini:1997eg}.

 Turning to the baryon asymmetry of the Universe, $\eta_B$, in
 Ref.~\cite{Semikoz:2009ye} it was noted that, due to the global
 charge conservation $$2{\rm d}L_{eL}/{\rm d}t=-{\rm d}L_{eR}/{\rm d}t
 + \dot{B}/3$$ and the presence of the Abelian anomaly term
 (\ref{Abelian}), at the electroweak phase transition epoch $\eta_B$
 "sits" in the hypermagnetic field~\cite{Giovannini:1997eg},
\begin{equation}\label{baryon}
\eta_B(t_{EW})=\frac{3g^{'2}}{4\pi^2s}\int_{t_0}^{t_{EW}}\left[\alpha_YB_Y^2 -\eta_Y(\nabla\times {\bf B}_Y)\cdot{\bf B}_Y\right]{\rm d}t,
\end{equation} 
where $s=2\pi^2g^*T^3/45$ is the entropy density.  Now substituting
$\alpha_Y$ for right-handed electrons, instead of that for neutrinos,
and neglecting the diffusion term $\sim \eta_Y$ one notes that today's
observed baryon asymmetry $$\eta_B\simeq 10^{-10}$$ is reproduced if
\begin{equation}\label{product}
  \left(\frac{\mid\xi_{eR}\mid}{0.0001}\right)\frac{B_Y^2(t_{EW})}{T_{EW}^4} \simeq 7\times 10^{-14},
\end{equation}
at the electroweak phase transition epoch. For example if we take
$\mid\xi_{eR}\mid\leq 10^{-5}$ we get from Eq.~(\ref{product})
$B_0\geq 10^{18}~{\rm G}$ for the initial Maxwellian seed field
$B_0=\cos\theta_WB_Y$ at $t_{EW}$.

Let us discuss our reference value choice for the right-electron
asymmetry $\mid\xi_{eR}\mid\simeq 0.0001$ used above. In the adiabatic
approximation $\dot {s}=\dot{T}=0$ for the right-electron asymmetry
density $n_{eR}=\mu_{eR}T^2/6$ one gets, from eq. (\ref{Abelian}), 
$$
\frac{{\rm d}\xi_{eR}}{{\rm dt}}=-\frac{6g^{'2}}{4\pi^2T^3}{\bf E}_Y{\bf B}_Y ,
$$
where  ${\bf E}_Y$, 
\footnote{Such field comes
from the Ohm law ${\bf J}=\sigma_{cond}[{\bf E}_Y + {\bf V}\times {\bf B}_Y]$, substituted into the 
generalized Maxwell equation $-\partial_t{\bf E}_Y + \nabla\times {\bf B}_Y={\bf J} + {\bf J}^Y_5$,
derived from the effective Lagrangian (\ref{Lagrangian2}) with the pseudovector current
${\bf J}^Y_5$ given by eq. (\ref{J5})
when neglecting in MHD approach the displacement current $\partial_t{\bf E}_Y$.}
$$
{\bf E}_Y=\eta_Y\nabla\times {\bf B}_Y - {\bf V}\times {\bf B}_Y -\alpha_Y{\bf B}_Y,
$$
is the hyper-electric field. With this one gets the ordinary
differential equation
\begin{equation}\label{diffur}
\frac{{\rm d}\xi_{eR}}{{\rm dt}} + [P(t) + \Gamma (t)]\xi_{eR}=Q(t).
\end{equation}
Here the coefficients $P, Q$ given by
$$
P(t)=\left(\frac{6g^{'2}}{4\pi^2T^3\sigma_{cond}}\right)\frac{g^{'2}B_Y^2(t)T}{4\pi^2},$$
$$Q(t)=-\left(\frac{6g^{'2}}{4\pi^2T^3\sigma_{cond}}\right)k_0B_Y^2(t)
$$
depend on $\xi_{eR}(t)$ through the hypermagnetic field amplitude in
eq. (\ref{arbitrary}).  In getting the last coefficient $Q$ we used the
Chern-Simons wave $Y_0=Y_z=0, ~~Y_x=Y_0(t)\sin k_0z,Y_y=Y_x=Y_0(t)\cos
k_0z$ as the simplest configuration for the hypercharge field which
allows to get $(\nabla\times {\bf B}_Y)\cdot {\bf B}_Y=k_0B_Y^2(t)$.
{In Eq.~(\ref{diffur}) $\Gamma=2\Gamma_{RL}$ denotes the rate for
  chirality flip processes (inverse Higgs decays $e_R\bar{e}_L\to
  \varphi^{(0)}$, $e_R\bar{\nu}_{eL}\to \varphi^{(-)}$ with
  equivalent rates), and we neglect for simplicity left lepton
  asymmetries.}

The formal solution of the nonlinear integro-differential equation
(\ref{diffur}) takes the form
$$
\xi_{eR}(t)=e^{-\int_{t_0}^t(P + 
\Gamma)dt}\left[\xi_{eR}^{(0)} + \int_{t_0}^tQ(t^{'})e^{\int_{t_0}^{t^{'}}[P(t^{''})+ \Gamma (t^{''})]dt^{''}}dt^{'}\right].
$$
Here we give the asymptotic solution for slowly changing hypermagnetic
fields when $P\approx const$, $Q\approx const$ assuming also
  $\Gamma\approx const$
\begin{equation}\label{solution}
\xi_{eR}(t)=\left[\xi_{eR}^{(0)} - \frac{Q}{P + \Gamma}\right]e^{-[P + \Gamma](t-t_0)} + \frac{Q}{P + \Gamma}\approx \frac{Q}{P + \Gamma}\approx -\frac{4\pi^2k_0}{g^{'2}T}.
\end{equation}
{In the last step in Eq. (\ref{solution}) we have neglected
  chirality flip rates in strong hypermagnetic field, $\Gamma\ll
  P$. Indeed one has~\cite{Campbell:1992jd},
\begin{equation}\label{gamma}
\Gamma_{RL}=0.88\times 10^{-14}\left[1 - \left(\frac{T_{EW}}{T}\right)^2\right]T,
\end{equation}
which is much less than the coefficient $P\sim B_Y^2$ above,
$$
P=5.6\times 10^{-7}\left[\frac{B_Y}{T^2}\right]^2 T.
$$
Note that for hypermagnetic field $B_Y(t)$ frozen-in ideal plasma the
ratio $B_Y/T^2=const$ during the cooling of the Universe.  Hence for
hypermagnetic field values, say, $B_Y=10^{-3}T^2$, our approximation
$P\gg \Gamma$ holds} \footnote{Note that this inequality holds
also for moderate fields at $T\sim T_{EW}$ where the rate
(\ref{gamma}) vanishes. However, such rate can modify the evolution of
$\xi_{eR}(t)$ above $T_{EW}$ and below chirality flip temperature
$T_{RL}$, $T_{RL}>T >T_{EW}$, when left leptons enter equilibrium with
the right ones, see below in Section 4.}.

One can easily check that the exponential term in (\ref{solution})
vanishes at the EWPT time, since $Pt_{EW}\gg 1$ for $B_Y<T^2$.  Then
taking into account for the survival condition of the Chern-Simons
wave versus ohmic diffusion, $k_0<10^{-7}T$, substituting weak
coupling $g^{'2}=0.12$ we get the estimate of the lepton asymmetry in
a strong hypermagnetic field,
$$
\mid\xi_{eR}\mid\leq 3.3\times 10^{-5},
$$
close to what we used above. This value can be enhanced for a more
complicated configuration of a helical hyperharge field accounting for
the linkage integer number $n$ (number of knots, $n=\pm 1, \pm 2,...$)
entering the coefficient $Q\sim {\bf B}_Y\cdot(\nabla\times {\bf
  B}_Y\sim n)$.


{The subsequent evolution of the initial Maxwellian field, $B_0=B_Y\cos
\theta _W$ given by Eq.~(\ref{product}) was studied in
Ref.~\cite{Semikoz:2009ye} and illustrated in Fig.~1 of that paper.}
Although the ``slope'' of hypermagnetic field enhancement found here
differs from that obtained in Ref.~\cite{Semikoz:2007ti} we note that
this does not alter in any significant way the
``magneto-baryogenesis'' mechanism proposed in
Ref.~\cite{Semikoz:2009ye}. A very similar generic evolution picture
emerges. Of course the causal approach we adopt cannot provide a
complete scenario for generating the proto-galactic large-scale fields
observed in astronomy. One needs to assume the presence of a
large-scale field somewhere after the electroweak phase transition.
For example a noncausal spectator field model, 
or a similar $\alpha$-helicity
mechanism induced by a nonvanishing neutrino asymmetry as in
Ref.~\cite{Semikoz:2003qt}.
In both cases the magnetic field scale can exceed the horizon size at
temperatures $T\leq T_{EW}$, $\Lambda_B\geq l_H(T)$, which later
enters the horizon and becomes frozen-in plasma decreasing in
amplitude in the standard way.  Simultaneously its scale increases
slower than the horizon, let us say, starting from the temperature
$T\leq 1~{\rm GeV}$, $z=10^{13}$, its evolution is given by
$\Lambda_B(z)=l_H(1~{\rm GeV})[(1+z)/(10^{13})]^{-1}<l_H\sim z^{-2}$.

As a result at the stage of a galaxy formation $z\simeq 10$ the
magnetic field scale is too small, only $\Lambda_B\simeq 1.44\times
10^{16}~cm\ll 1~pc$ . Thus, there are $N=(1~pc/\Lambda_B)^3=8\times
10^6$ expanding domains of randomly oriented fields for which the mean
field $B_{mean}=B(z=10)/\sqrt{N}$ has the large initial galactic scale
$\Lambda=1~pc$ and a small amplitude of the order $B_{mean}\simeq
3.5\times 10^{-13}~G$. The corresponding dynamo mechanism then
provides amplification up to $B_{gal}\sim 10^{-6}~G$.
This way we reconfirm the viability of a unified scenario proposed in
Ref.~\cite{Semikoz:2009ye}.

 Let us comment on the difficulty to amplify large-scale hypermagnetic
fields in any causal scenario.  This holds in our dynamo mechanism as
well as for the case where additional pseudoscalar coupling with
hypercharge fields are assumed as in \cite{Brustein:1999rk}.
Therefore a preliminary amplification (of a seed field $B_0^Y$) like
in the inflationary scenario seems to be important for large-scale fields.

\section{Leptogenesis at $T_{EW}<T<T_{RL}$}

We now turn to the issue of the possible washout of the electron
asymmetry in the electroweak plasma in the early universe at
temperatures below the chirality flip temperature $T_{RL}$.
The evolution of the asymmetries is determined by processes involving
the Higgs scalar, as well as a pseudoscalar term proportional to
$\mathbf{E}_Y\mathbf{B}_Y$ and related to the anomaly.

{In Quantum Electrodynamics} the electron lepton number is conserved and
the Abelian anomaly terms in an external electromagnetic field
$F_{\mu\nu}$~\cite{Zee} \footnote{We change here sign of $\gamma_5$
  matrix, $\gamma_5\to -\gamma_5$, with respect to the notation used
  in \cite{Zee}. This is in the agreement with the notation and sign
  for Abelian anomaly in Ref.~\cite{Giovannini:1997eg}.},
\begin{equation}
  \frac{\partial j^{\mu }_L}{\partial x^{\mu}} =
  +\frac{e^2}{16\pi^2}F_{\mu\nu}\tilde{F}^{\mu\nu},
  \quad
  \frac{\partial j^{\mu }_R}{\partial x^{\mu}} =
  -\frac{e^2}{16\pi^2}F_{\mu\nu}\tilde{F}^{\mu\nu},
\end{equation}
do not contradict to this law,
$\partial_{\mu}j^{\mu}=\partial_{\mu}(j_L^{\mu} + j_R^{\mu})=0$.
{Conversely}, in the present case, $\partial_{\mu}(j_L^{\mu} + j_R^{\mu})\neq 0$,
since the coupling constants $e\to g^{'}y_{R,L}/2$ are different for
right singlet $e_R$ and left doublet $L=(\nu_{eL}e_L)^T$. In other
words, the coupling of the hypercharge field is chiral, while
Maxwellian field has a vector-like coupling to fermions.
This difference ensures a non-zero electron asymmetry in the
electroweak plasma also below $T_{RL}$.

Note that in the kinetic equations for lepton asymmetries we must
include the Abelian anomalies for right electrons (\ref{Abelian}) as
well as for the left doublet:
\begin{equation}\label{left}
  \frac{\partial j^{\mu}_{L}}{\partial x^{\mu}} =
  \frac{g^{'2}y_L^2}{64\pi^2}Y_{\mu\nu}\tilde{Y}^{\mu\nu} =
  \frac{g^{'2}}{16\pi^2}\mathbf{E}_Y\mathbf{B}_Y,
  \quad
  y_L=-1.
\end{equation}

Here we are concerned with the stage of chirality flip processes when
left electrons and left neutrinos enter the equilibrium with the right
electrons through the Higgs decays (inverse decays).  These left
leptons have the same densities at temperatures below $T<T_{RL}$,
$n_L=n_{eL}=n_{\nu_{eL}}$, and the same chemical potentials
$\mu_{eL}=\mu_{\nu_{eL}}$.

Taking into account the Abelian anomalies for $e_R$ and left doublet
$L=(\nu_{eL}e_L)^T$ one gets the following system of kinetic equations
for the lepton asymmetry densities $n_R=n_{eR}-n_{\bar{e}R}$,
$n_L=n_{eL}-n_{\bar{e}L}$, or for the corresponding lepton numbers
$L_{eL}=L_{\nu_{eL}}=n_L/s$, $L_{eR}=n_R/s$:
\begin{eqnarray}\label{system}
  &&\frac{{\rm d}L_{eR}}{\rm dt} =-
  \frac{g^{'2}}{4\pi^2s}\mathbf{E}_Y\mathbf{B}_Y + 2\Gamma_{RL}(L_{eL}-L_{eR}),
  \quad
  {\rm for}
  \quad
  e_R\bar{e}_L\to \varphi^{(0)}, e_R\bar{\nu}_{eL}\to \varphi^{(-)},\nonumber\\
  &&\frac{{\rm d}L_{eL}}{\rm dt} =
  \frac{g^{'2}}{16\pi^2s}\mathbf{E}_Y\mathbf{B}_Y +\Gamma_{RL}(L_{eR} - L_{eL}),
  \quad
  {\rm for}
  \quad
  \bar{e}_Re_L\to \bar{\varphi}^{(0)},\nonumber\\&&\frac{{\rm d}L_{\nu_{L}}}{\rm dt} =
  \frac{g^{'2}}{16\pi^2s}\mathbf{E}_Y\mathbf{B}_Y +\Gamma_{RL}(L_{eR} - L_{eL}),
  \quad
  {\rm  for}
  \quad
  \bar{e}_R\nu_{eL}\to \varphi^{(+)}.
\end{eqnarray}
Here the factor 2 in the first line takes into account the equivalent
reaction branches for inverse Higgs scalar decays and for simplicity
we neglected Higgs boson asymmetries, $n_{\varphi}=n_{\bar{\varphi}}$,
so Higgs decays into leptons do not contribute in the kinetic
equations~\eqref{system}.

Summing the equations~\eqref{system} one can easily see that the
inverse Higgs processes $\sim \Gamma_{RL}$ do not contribute in the
Hooft's rule $d\eta_B/dt=3[dL_{eR}/dt + dL_{eL}/dt +
dL_{\nu_{eL}}/dt]$ directly. In other words, we find that leptogenesis
exists at the stage $T<T_{RL}$ and the baryon asymmetry is generated
through hypermagnetic fields.

Note that for $T>T_{RL}$, before left leptons enter equilibrium with
right electrons, the anomaly (\ref{left}) was not efficient since the left
electron (neutrino) asymmetry was zero, $\mu_{eL}=\mu_{\nu_{eL}}=0$,
while a non-zero primordial right electron asymmetry, $\mu_{eR}\neq
0$, kept the baryon asymmetry at the necessary level.
In other words, for $T>T_{RL}$ the anomaly (\ref{left}) was present at the
stochastic level, with $<\delta j_L^{\mu}>=0=<{\bf E}_Y{\bf B}_Y>$
valid only on large scales.

{We stress that $L_{eL}(T_0)=0=\mu_{eL}$ initially, at
  $T_0=T_{RL}$, while $L_{eR}$ grew before $T_{RL}$.  Therefore
  $L_{eR} - L_{eL} >0$ remains non-zero till $T_{EW}$. We have checked
  quantitatively that this inequality holds as a result of the kinetic
  equations, though a detailed study of this issue is beyond scope of
  the present work. In contrast, in Ref.~\cite{Giovannini:1997eg}
  $\mu_{eL}=0$ below $T_{RL}$ all the way down to $T_{EW}$, while we
  assume this only as an initial condition, $\mu_{eL}(T_0)=0$ valid at
  $T_0=T_{RL}$ and take into account the corresponding chirality flip
  reactions associated to Higgs boson decays (inverse decays). 
  As a result, the wash-out of the lepton and baryon asymmetries at
  $T<T_{RL}$ does not occur in our scenario.}

\section{Summary}

In short, by statistically averaging the standard model effective
Lagrangian in the presence of a  seed hypermagnetic field
we have provided a novel physical interpretation of the parity
violating Chern-Simons term as a polarization effect associated to
massless right-handed electrons (positrons) moving in the plasma.
The hyper-magnetic field will grow exponentially via the dynamo effect
and may induce today's observed cosmological baryon asymmetry,
re-confirming the viability of a unified scenario proposed in
Ref.~\cite{Semikoz:2009ye} relating the cosmological baryon asymmetry
with the origin of the protogalactic large-scale magnetic fields
observed in astronomy.

As a final comment we mention the existence of alternative ways to
connect the baryon asymmetry with hypermagnetic fields, as suggested
in Ref.~\cite{Bamba:2007hf} using physics beyond the Standard Model.
\acknowledgments We thank M. Shaposhnikov and V. Rubakov for useful
discussions.  This work was supported by the Spanish MICINN under
grants FPA2008-00319/FPA, FPA2011-22975, and MULTIDARK CSD2009-00064
(Consolider-Ingenio 2010 Programme), by Prometeo/2009/091 (Generalitat
Valenciana), by the EU Network grant UNILHC PITN-GA-2009-237920.

 \bibliographystyle{h-physrev4}

\end{document}